\documentclass[prb,twocolumn,superscriptaddress,showpacs,preprintnumbers,amsmath,amssymb,floats]
{revtex4}

\usepackage{txfonts}
\usepackage{amssymb}
\usepackage{graphicx}
\usepackage{color}

\begin{document}

\title{Observation of the anisotropic Dirac cone in the band dispersion of 112-structured iron-based superconductor Ca$_{0.9}$La$_{0.1}$FeAs$_{2}$}

\author{Z. T. Liu}
\affiliation{State Key Laboratory of Functional Materials for Informatics,
Shanghai Institute of Microsystem and Information Technology (SIMIT),
Chinese Academy of Sciences, Shanghai 200050, China}
\author{X. Z. Xing}
\affiliation{Department of Physics and Key Laboratory of MEMS of the Ministry of Education,
Southeast University, Nanjing 211189, China}
\author{M. Y. Li}
\affiliation{State Key Laboratory of Functional Materials for Informatics,
Shanghai Institute of Microsystem and Information Technology (SIMIT),
Chinese Academy of Sciences, Shanghai 200050, China}
\author{W. Zhou}
\affiliation{Department of Physics and Key Laboratory of MEMS of the Ministry of Education,
Southeast University, Nanjing 211189, China}
\author{Y. Sun}
\affiliation{Department of Physics and Key Laboratory of MEMS of the Ministry of Education,
Southeast University, Nanjing 211189, China}
\author{C. C. Fan}
\author{H. F. Yang}
\author{J. S. Liu}
\affiliation{State Key Laboratory of Functional Materials for Informatics,
Shanghai Institute of Microsystem and Information Technology (SIMIT),
Chinese Academy of Sciences, Shanghai 200050, China}
\author{Q. Yao}
\affiliation{State Key Laboratory of Functional Materials for Informatics,
Shanghai Institute of Microsystem and Information Technology (SIMIT),
Chinese Academy of Sciences, Shanghai 200050, China}
\affiliation{State Key Laboratory of Surface Physics, Department of Physics,
and Advanced Materials Laboratory, Fudan University, Shanghai 200433, China}
\author{W. Li}
\affiliation{State Key Laboratory of Functional Materials for Informatics,
Shanghai Institute of Microsystem and Information Technology (SIMIT),
Chinese Academy of Sciences, Shanghai 200050, China}
\affiliation{State Key Laboratory of Surface Physics, Department of Physics,
and Advanced Materials Laboratory, Fudan University, Shanghai 200433, China}
\affiliation{CAS-Shanghai Science Research Center, Shanghai 201203, China}
\author{Z. X. Shi}
\affiliation{Department of Physics and Key Laboratory of MEMS of the Ministry of Education,
Southeast University, Nanjing 211189, China}
\author{D. W. Shen}\email{dwshen@mail.sim.ac.cn}
\affiliation{State Key Laboratory of Functional Materials for Informatics,
Shanghai Institute of Microsystem and Information Technology (SIMIT),
Chinese Academy of Sciences, Shanghai 200050, China}
\affiliation{CAS-Shanghai Science Research Center, Shanghai 201203, China}
\affiliation{CAS Center for Excellence in Superconducting Electronics (CENSE), Shanghai 200050, China}
\author{Z. Wang}
\affiliation{State Key Laboratory of Functional Materials for Informatics,
Shanghai Institute of Microsystem and Information Technology (SIMIT),
Chinese Academy of Sciences, Shanghai 200050, China}

\begin{abstract}
    CaFeAs$_{2}$ is a parent compound of recently discovered 112-type iron-based superconductors. It is predicted to be a staggered intercalation compound that naturally integrates both quantum spin Hall insulating and superconducting layers and an ideal system for the realization of Majorana modes. We performed a systematical angle-resolved photoemission spectroscopy and first-principle calculation study of the slightly electron-doped CaFeAs$_{2}$. We found that the zigzag As chain of 112-type iron-based superconductors play a considerable role in the low-energy electronic structure, resulting in the characteristic Dirac-cone like band dispersion as the prediction. Our experimental results further confirm that these Dirac cones only exists around the $X$ but not $Y$ points in the Brillouin zone, breaking the $S_4$ symmetry at iron sites. Our findings present the compelling support to the theoretical prediction that the 112-type iron-based superconductors might host the topological nontrivial edge states. The slightly electron doped CaFeAs$_{2}$ would provide us a unique opportunity to realize and explore Majorana fermion physics.
\end{abstract}

\maketitle

Topological superconductors have attracted great research interests since they are proposed to host the unconventional Andreev surface states, which are a condensed-matter realization of hypothetical Majorana fermions originating from the field of particle physics~\cite{Majorana1,Majorana2,Majorana3,Majorana4,Majorana5,Majorana6}. Besides the novelty in the fundamental research, the experimental realization of Majorana fermions in solids is as well of significance for application since it is suggested to be a key to achieving the fault-tolerant topological quantum computing~\cite{computer1,computer2,computer3}. In this context, a diversity of candidate materials which are feasible to host Majorana fermions have been proposed in the past several years, including quantum wires in proximity to superconductors~\cite{Majorana4,a}, hybrid superconductor-semiconductor nanowires~\cite{Majorana4,b,A,B,C} and topological insulators coupled to superconductors~\cite{c,d,e,f}. Interestingly, all these proposals are restricted to utilize the coupling between topological insulators and ordinary superconductors. Recently, high-temperature (high-$T_c$) superconductors such as Bi$_2$Sr$_2$CaCu$_2$O$_{8+x}$ (Bi-2212) have been also applied as the substrate materials to synthesize artificial topological insulator/superconductor heterostructures, based on the expectation that a larger superconducting gap could be resolved since more robust cooper pairs might be introduced into the normal side through superconducting proximity effect~\cite{D,E,F}. However, whether this proximity system can really host Majorana fermions is still lack of further support, and whether high-$T_c$ superconductors could work in realizing Majorana fermions is a seductive but still open question~\cite{JPHu1,JPHu2}.

Very recently, there is one fascinating proposal which suggests the newly discovered 112-type iron-based high-temperature superconductor parent compound CaFeAs$_{2}$ be naturally a promising candidate to host Majorana fermions~\cite{JPHu2}. CaFeAs$_{2}$ has a structure with alternating CaAs and FeAs layers [see Fig. 1(a)]. While FeAs layers have been proven to be responsible for the high-$T_c$ superconductivity similar to other iron-based superconductors~\cite{FeSC1}, this proposal has shown that each CaAs layer is actually a topologically nontrivial two-dimensional quantum spin Hall insulator. Consequently, below $T_c$ the proximity effect would drive the zigzag As chain of the CaAs layer to become a one-dimensional topological superconductor. In this proposal, the key argument is that the $p_x$ and $p_y$ orbitals of the As atoms in the unique arsenic zigzag bond layers of CaFeAs$_{2}$ would result in an anisotropic Dirac cone near the Fermi level ($E_F$), which is gapped by the spin-orbit coupling (SOC) and topological nontrivial~\cite{JPHu1,JPHu2}. Once this Dirac-cone like band dispersion could be confirmed experimentally, the staggered intercalation of quantum spin Hall insulator and superconductor in this 112-type iron-based superconductor could provide us a unique opportunity to realize Majorana modes in solids. However, this theory is based on the 112-type iron-based superconductor parent CaFeAs$_{2}$, which actually has not been synthesized out experimentally yet. Thus, whether this prediction as well applies to the corresponding electron-doped superconductor Ca$_{1-x}$La$_{x}$FeAs$_{2}$ with a $T_c$ as high as 40 K would be an ideal touchstone for this intriguing model. Besides, so far there still have been no specific reports on this characteristic but tenuous feature in the band dispersion measurements of 112-type iron-based superconductors~\cite{ARPES1,ARPES2,ARPES3}.

In this letter, using both first-principle calculations and angle-resolved photoemission spectroscopy (ARPES), we resolved the characteristic Dirac-cone like band dispersion of this material and then performed a comprehensive investigation of this band feature. We find that the CaAs layers indeed play a considerable role in its low-energy electronic structure, As $4p$ orbitals of which form the unique Dirac cone-like band dispersion distinct from those of other iron-based superconductors. Our experimental results further confirm that these Dirac cones only exist around the $X$ but not $Y$ points in the Brillioun zone, breaking the $S_4$ symmetry at iron sites. In addition, these Dirac cones are rather two-dimensional nature, which show negligible photon energy dependence. Our findings present the compelling support to the theoretical prediction that the 112-type iron-based superconductors indeed host the topological nontrivial bands originating from the unique zigzag As chain.


\begin{figure}[t]
  \includegraphics[width=7.6cm]{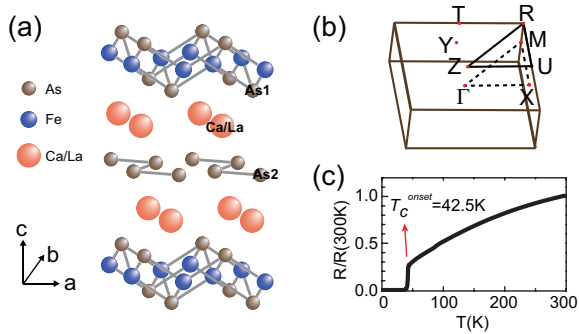}\\
  \caption{(a) Crystal structure of Ca$_{1-x}$La$_{x}$FeAs$_2$.(b) Three-dimensional Brillouin zone of Ca$_{1-x}$La$_{x}$FeAs$_2$. (c) Temperature dependence of normalized resistance in ab-plane of Ca$_{0.9}$La$_{0.1}$FeAs$_{2}$ single crystals.
  }\label{character}
\end{figure}

\begin{figure}[t]
  \includegraphics[width=7.5cm]{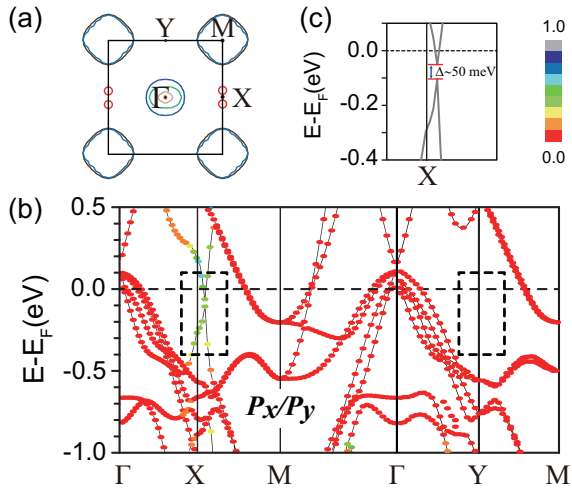}\\
  \caption{(a) The calculated Fermi surface. (b) The calculated orbital-projected band structure of Ca$_{0.9}$La$_{0.1}$FeAs$_{2}$. The weight of individual As 4\emph{p$_x$}(\emph{p$_y$}) orbitals for different bands is presented by gradually varied colors. The dashed boxes indicate the bands near $X$ and $Y$. (c) The zoom-in band structure near the $X$ point of BZ in the vicinity of the Fermi level.}\label{Dirac}
\end{figure}

High-quality single crystals with nominal composition Ca$_{0.9}$La$_{0.1}$FeAs$_{2}$ were synthesized using the flux method as descried in previous report~\cite{growth,growth1,growth2,growth3}. Fig. 1(a) illustrates the schematics of the crystal structure of Ca$_{1-x}$La$_{x}$FeAs$_{2}$, containing alternately stacked FeAs and CaAs layers as aforementioned. The three-dimensional reduced Brillouin zone (BZ), corresponding to the unit cell containing two Fe atoms, is shown in Fig. 1(b). The normalized resistivity in the $ab$ plane gives a $Tc$ onset of 42.5 K, as illustrated in Fig. 1(c).

Our ARPES measurements were conducted at the beamline I05 of the Diamond Light Source (DLS), and the beamline 13U of the National Synchrotron Radiation Laboratory (NSRL). All the data were taken at 10 K in an ultrahigh vacuum better than 8$\times$10$^{-11}$ torr. This setup is equipped with a VG-Scienta R4000 electron analyzer. The angular and the energy resolution was set to 0.2$^{\circ}$ and 10-20 meV, respectively. The electronic band calculations were performed within the density-functional formalism as implemented in the VASP code~\cite{VASP}. The plane-wave-basis method and the Perdew-Burke-Ernzerh of exchange-correlation potential were applied.

\begin{figure*}[t]

\includegraphics[width=14.5cm]{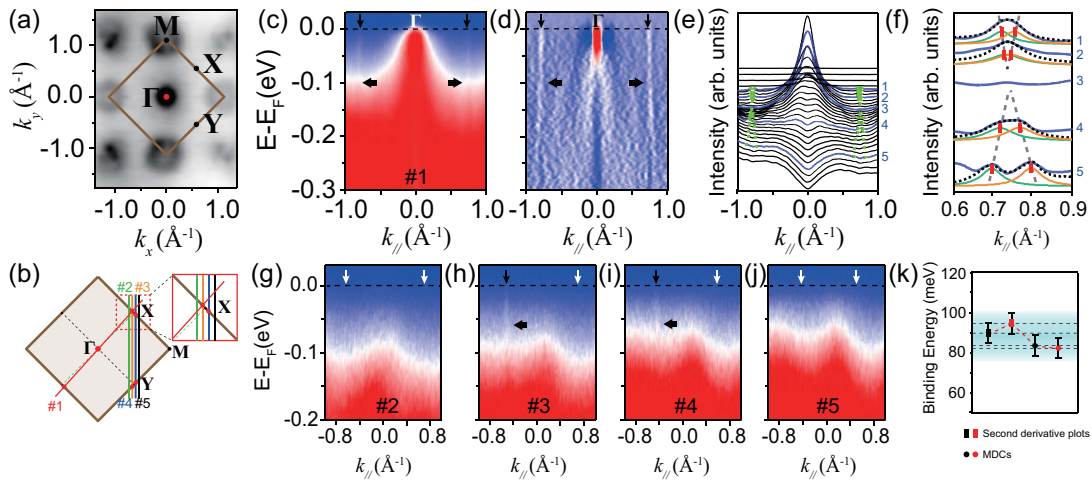}\\
  \caption{(a) Photoemission integrated intensity map at $E_F$ ($\pm$10 meV) of Ca$_{0.9}$La$_{0.1}$FeAs$_{2}$ using 105 eV photons. The brown lines indicate the BZ. (b) The indication of the cut direction in the projected two-dimensional BZ. (c)-(e) The photoemission intensity plots and the corresponding momentum distribution curves (MDCs) along Cut $\#$1 taken with 105 eV photons. (f) Lorentzian peaks fitting to a few representative  MDCs (Nos. 1-5). (g)-(j) Photoemission intensity plots along Cuts $\#$2$\sim$5 taken with 105 eV photons. The arrows are guides to the eye for the momenta where the Dirac cone-like band dispersions can be detected. (k) Energy location of the Dirac point with error bars determined from the second derivative plots and MDCs.}\label{kz}
\end{figure*}

We calculated the low-energy electronic structure of Ca$_{0.9}$La$_{0.1}$FeAs$_{2}$, in which the chemical potential was determined by best fitting the experimental Fermi surface (FS) [Fig. 3(a)] to the DFT calculation and the SOC was applied. Similar to other iron-based superconductors, the calculated FS of this sample [Fig. 2(a)] shows the common three hole-like pockets around BZ center and two electron pockets around the corner attributed to the Fe $3d$ orbitals~\cite{ARPES3}. Besides, we note that there also exist the extra small red Fermi pockets around $X$. Both the corresponding spaghetti drawing of the band structure [Fig. 2(b)] and its close-up [Fig. 2(c)] around $X$ clearly show that these small Fermi pockets are composed of Dirac-cone like bands around the $X$ point, which is in good agreement with the proposal for CaFeAs$_{2}$~\cite{JPHu1,JPHu2}. Particularly, the gap isolating the upper and lower Dirac band still remains though the carrier dosage has shifted the chemical potential away from the Dirac node as in CaFeAs$_{2}$. However, the gap size in Ca$_{0.9}$La$_{0.1}$FeAs$_{2}$ is only $\sim$50 meV, smaller than that predicted for CaFeAs$_{2}$ (100 meV). Furthermore, we have extracted the orbital characters of low-lying bands of Ca$_{0.9}$La$_{0.1}$FeAs$_{2}$, and the weight of individual As $4p_x$($4p_y$) orbitals of the zigzag As chain for different bands is presented by the gradually varied colors. It demonstrates that the Dirac-cone like bands near $X$ point are mainly contributed by the two-dimensional 4$p_x$($p_y$) orbitals of the zigzag As in Ca$_{0.9}$La$_{0.1}$FeAs$_{2}$ or CaFeAs$_2$, which are suggested to play vital roles in determining the topological nontrivial properties of CaAs layers in the theoretical proposal.

Fig. 3(a) shows the photoemission intensity map, which agrees well with the DFT calculation result. However, it is difficult to resolve the predicted small pockets around $X$ induced by the Dirac-cone like bands straightforwardly from the map. This might be due to the overwhelming intensity of some broad bands from Fe $3d$ orbitals. To seek for the tenuous Dirac cones in the vicinity of $E_F$, we performed a detailed survey of the photoemission spectra around both $X$ and $Y$ points in the BZ, as illustrated in Fig. 3(b). In Fig. 3(c), the intensity plot shows the band dispersion taken along the Cut$\#1$, slightly off the $\Gamma-X$ direction but crossing the Dirac cone around $X$. This data together with the corresponding second derivative plot [Fig. 3(d)] and momentum distribution curves (MDCs) [Fig. 3(e)] clearly show fast dispersing bands around $X$ with the characteristic linearity, which is  quantitatively confirmed by the peaks fitting of MDCs as shown in Fig. 3(f). Also, the Dirac point energy and Dirac Fermi velocity are determined to be 90$\pm$10 meV below $E_F$ [Fig. 3(k)] and (3.8$\pm$0.3)$\times$10$^{5}$ m/s, respectively, which are in remarkable agreement with the theoretically predicted values 80 meV below $E_F$ and 4.18$\times$10$^{5}$ m/s.

\begin{figure*}[t]
  \includegraphics[width=15cm]{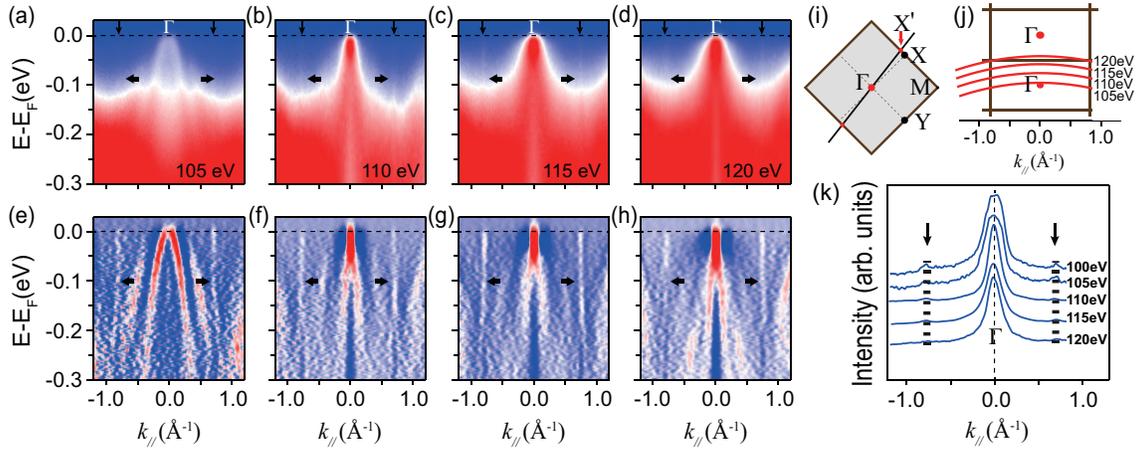}\\
  \caption{Photo-energy dependence of Dirac cone near the X point of BZ. (a)-(d) The photoemission intensity plots along $\Gamma-X'$ taken with different photo-energy. (e)-(h) The second derivative plots corresponding to (a)-(d) The lines and arrows are guides to the eye for the Dirac cone-like band dispersions. (i) The indication of the cut direction in the projected two-dimensional BZ. (k) The photon-energy dependence of the MDCs near E$_F$ along $\Gamma-X'$.
  }\label{character}
\end{figure*}

For the Ca$_{1-x}$La$_{x}$FeAs$_{2}$, the As atoms in the CaAs layers form zigzag chains with a distorted checkerboard lattice in which an As atom moves in the $x$ direction and is off the center. Thus, the $S_4$ symmetry for As $4p$ orbitals should be broken. The $X$ and $Y$ points would no longer be equivalent in BZ, and thus the Dirac-cone like feature could only be observed around one point~\cite{JPHu1,JPHu2}. Actually, our DFT calculation [Fig. 2(b)] and detailed spectral scans around both $X$ and $Y$ provide the compelling proof of this prediction. While the photoemission data along Cut \#1 can reveal Dirac bands around both equivalent X points [Figs. 3(c-d)], the data along Cuts \#2$\sim$5 do not show any sign of photoemission spectral weight around the $Y$ point. Here, a detailed survey around $Y$ was performed to prevent missing the small momentum region in which the Dirac cones could exist. Note that the photoemission experimental geometry for Cuts \#1 and \#2$\sim$5 is exactly the same except of the different mounting azimuthal angles of samples [Fig. 3(b)], so the interference of matrix element effect should be minimized. Consequently, this result clearly confirms the anisotropy of these Dirac cone-like bands.

To comprehensively understand these Dirac cones in the three-dimensional BZ, we have performed detailed $k_z$ dependent measurement of the bands around the $X$ point taken along the $\Gamma(Z)-X'(R')$ direction [Fig 4(i)]. Using the free-electron final-state model~\cite{Kz}, in which the inner potential was set to 15 eV~\cite{ARPES3}, we can estimate the $k_z$ value probed with different photons. Figs. 4(a-d) show the photoemission intensity plots taken with photon energies ranged from 105 eV to 120 eV. Together with the corresponding second derivative plots [Figs. 4(e-h)], we find some hole-like bands around the $\Gamma(Z)$ points demonstrate substantial photon energy dependence, which is in consistent with our previous report on the considerable three-dimensionality of bands attributed to Fe $3d$ orbitals in Ca$_{1-x}$La$_{x}$FeAs$_{2}$~\cite{ARPES3}. However, for the Dirac cones around $X(R)$, their band dispersions show negligible $k_z$ dependence, as summarized in the MDCs [Fig. 4(k)], indicating their rather two-dimensional nature considering the fact that the photon energies we applied have covered more than half of the BZ along $k$$\perp$ [Fig. 4(j)]. This finding further confirms that the Dirac-cone like bands and the in-gap states around $X$ are attributed to the more two-dimensional As $4p_x$($p_y$) orbitals in the zigzag chain.


Our DFT calculation and experimental data confirm that the unique As zigzag chain in 112-type iron-based superconductors would result in anisotropic Dirac cones in band dispersion as the theoretical prediction. For such a material like CaFeAs$_{2}$, the calculation of the Wannier function centers of the occupied has shown that the Wannier states would switch partners and there should exist topologically nontrivial spin-polarized edge states in the gap~\cite{JPHu2}. On the other hand, although our photoemission experiments do not directly reveal the superconducting gap of this compound below $T_c$ due to the sample quality, previous ARPES result has implied the s-wave superconductivity in 112-type and other iron-based superconductors~\cite{ARPES3,FeSC2}. Thus, both essential ingredients of hosting topological superconductivity should have been fulfilled in lightly electron doped CaFeAs$_{2}$, and this compound is an ideal system for the realization of Majorana modes.

However, although we observe some weak spectral weight in between the upper and lower Dirac bands, we can still not pin down these in-gap states as the long-sought topological non-trivial edge states due to their extremely weak and blurred feature. Particularly, the characteristic Dirac point of the edge states inside the bulk valence band gap cannot be resolved. This might be still limited by the single crystal quality in the current stage or the rather weak spectral weight of the one-dimensional edge states. These edge states might be checked by more local and surface sensitive probes such as scanning tunneling microscopy in the future. Besides, to realize topological superconductors, an appropriate amount of electron doping to as-grown CaFeAs$_{2}$ is needed so as to induce necessary carriers to the FeAs layers and shift the chemical potential to cross the down edge bands only.


In summary, we report a systematical DFT calculation and ARPES study of the low-lying electronic structure of the the 112-type iron-based superconductor Ca$_{0.9}$La$_{0.1}$FeAs$_{2}$. Besides the commonly hole- and electron-like bands as reported in other iron-based pnictides, there exists extra Dirac-cone like bands around the $X$ point of the BZ, which are rather anisotropic and two-dimensional. We figure out a rather promising candidate Ca$_{0.9}$La$_{0.1}$FeAs$_{2}$ of the long-sought topological superconductors, and provide compelling evidences. Our work may provide a promising platform for researchers to explore Majorana mode which is the key for the realization of topological quantum computer.


We gratefully acknowledge the helpful discussions with Dr. Meixiao Wang. This work is supported by the National Key R\&D Program of the MOST of China (Grant No. 2016YFA0300204) and the National Science Foundation of China (Grant Nos. 11274332, 11574337, 11227902 and U1432135). D. W. S. is also supported by the ``Strategic Priority Research Program (B)'' of the Chinese Academy of Sciences (Grant No. XDB04040300) and the ``Youth Innovation Promotion Association CAS''.

\end{document}